# Efficient Implementation of an Accurate Algebraic Scheme for Sharp Interface Advection in Multiphase Flows

**Mehran Sharifi***

*Department of Mechanical Engineering, Amirkabir University of Technology (Tehran Polytechnic), Iran*

*****Corresponding Author**
Mehran Sharifi, Department of Mechanical Engineering, Amirkabir University of Technology (Tehran Polytechnic), Iran.

## Abstract
This study presents an efficient algebraic scheme known as MULES for sharp interface advection, verified against various schemes including first-order upwind, second-order central, van Leer flux limiter, and Geometric Volume-of-Fluid (VOF). Two problems involving a droplet in a two-dimensional (2D) vortex and a stationary droplet were examined. The model assessed the effects of the Interface Compression (IC) coefficient, ranging from 0 to 2, analyzing parameters such as Interface Advection Error (IAE) and Mass Conservation Error (MCE). Results indicated that increasing IC values enhanced interface tracking accuracy but introduced non-physical instabilities at higher values, compromising mass conservation. Specifically, the IAE decreased from 4.8% to 3.95% as IC increased from 0 to 2, showing a favorable effect until IC surpassed 1.4, where IAE fluctuated around 4%. Conversely, the MCE rose steeply from 0% to 23.19%, driven by parasitic currents and numerical instabilities. Additionally, MULES and van Leer flux limiter schemes evaluated volume fraction smoothing effects. Initial filtering reduced Dimensionless Pressure Difference (DPD) and Capillary Number (Ca), stabilizing the solution, but excessive filtering reintroduced numerical errors and instabilities. With one filtering step, DPD reduced by 0.23 and Ca dropped significantly by 73.31%, improving solution stability. However, further filtering increased DPD and Ca, reflecting the reintroduction of numerical errors. The maximum velocity of parasitic flow around the droplet initially decreased by almost 75% but increased by 30.92% with excessive filtering. IAE increased from 0.7 to 0.9 with initial filtering, then decreased to 0.63 with additional steps, indicating improved solver performance on smoother interfaces.

**Keywords:** Sharp Interface Advection, Multiphase Flows, Numerical Instabilities, Parasitic Currents, Algebraic Scheme

## Nomenclature

**Latin symbols**

| | | |
|---|---|---|
| $A$ | : | Anti-diffusive flux, $m^2\,s^{-1}$ |
| $D$ | : | Droplet diameter, $m$ |
| ET | : | Elapsed Time, $s$ |
| $F_c$ | : | Corrected flux, $m^3\,s^{-1}$ |
| $F_f$ | : | Advective flux on the cell face, $m^3\,s^{-1}$ |
| $F_h$ | : | Central-differencing flux, $m^3\,s^{-1}$ |
| $F_u$ | : | First-order upwind flux, $m^3\,s^{-1}$ |
| $\mathbf{F}_\sigma$ | : | Surface tension vector, $N\,m^{-3}$ |
| $g$ | : | Magnitude of gravitational acceleration, $m\,s^{-2}$ |
| $l$ | : | Domain length, $m$ |
| $m_x$ | : | Discretized form of volume fraction gradient in X-direction, – |
| $m_y$ | : | Discretized form of volume fraction gradient in Y-direction, – |
| $\mathbf{n}$ | : | Unit normal vector to the interface, – |
| $N_x$ | : | Number of cells in X-direction, – |
| $N_y$ | : | Number of cells in Y-direction, – |
| $p$ | : | Pressure, $Pa$ |
| $P$ | : | Cell anti-diffusive flux, $m^2\,s^{-1}$ |
| $Q_{i,j}$ | : | Total transported flux, $m^3\,s^{-1}$ |
| $r_f$ | : | Successive gradients, – |
| $\mathbf{S}$ | : | Surface area vector, $m^2$ |
| $t$ | : | Time, $s$ |
| $T$ | : | Periodicity, $s$ |
| $u$ | : | X-component of velocity, $m\,s^{-1}$ |
| $\mathbf{U}$ | : | Velocity field vector, $m\,s^{-1}$ |
| $v$ | : | Y-component of velocity, $m\,s^{-1}$ |
| $V_{i,j}$ | : | Cell volume, $m^3$ |
| $x_c$ | : | X-component of droplet's position, $m$ |
| $y_c$ | : | Y-component of droplet's position, $m$ |

**Greek symbols**

| | | |
|---|---|---|
| $\alpha$ | : | Volume fraction, – |
| $\delta_s$ | : | Delta function, – |
| $\eta$ | : | Discretized form of the unit normal vector to the interface, – |
| $\kappa$ | : | Mean radius of interface curvature, $m^{-1}$ |
| $\lambda_m$ | : | MULES limiter, – |
| $\mu$ | : | Dynamic viscosity, $Pa\,s$ |
| $\xi$ | : | Step function, – |
| $\rho$ | : | Density, $kg\,m^{-3}$ |
| $\sigma$ | : | Surface tension coefficient, $N\,m^{-1}$ |
| $\phi_f$ | : | Volume flux, – |
| $\phi_{rf}$ | : | Corrected volume flux, – |
| $\psi$ | : | Stream function, $m^2\,s^{-1}$ |
| $\omega_f$ | : | Van Leer's flux limiter function, – |
| $\omega_{rf}$ | : | Flux limiter of the IC scheme, – |

**Subscripts**

| | | |
|---|---|---|
| 0 | : | Initial value |
| D | : | First downwind cell |
| f | : | Cell face value |
| max | : | Maximum value |
| N | : | Neighbor cell |
| p | : | Primary phase (surroundings) |
| P | : | Current cell |
| q | : | Secondary phase (droplet) |
| r | : | Relative value |
| U | : | First upwind cell |
| UU | : | Second upwind cell |

**Superscripts**

| | | |
|---|---|---|
| ~ | : | Filtered value |
| 0 | : | Initial state |
| $v$ | : | Iteration number |
| min | : | Minimum value |
| max | : | Maximum value |
| $n$ | : | Current time step |
| $T$ | : | Final state |

**Dimensionless parameters**

| | | |
|---|---|---|
| Ar | : | Archimedes Number |
| Ca | : | Capillary Number |
| DPD | : | Dimensionless Pressure Difference |
| Eo | : | Eotvos Number |
| IAE | : | Interface Advection Error |
| IC | : | Interface Compression Coefficient |
| MCE | : | Mass Conservation Error |
| Mo | : | Morton Number |

## 1. Introduction

Two-phase flows, a term commonly used in computational modeling, pertain to issues involving two fluid phases. On the other hand, multiphase flows encompass a broader spectrum, extending to particle-laden flows. Despite analytical studies like Plateau on such flows dating back to the 19th century, the scope of analytical work is often significantly limited, even for relatively straightforward problems [1]. Experimental observations for practical applications pose challenges due to the difficulty of adapting experimental techniques to multiphase flows [2]. Consequently, there arises a necessity for the development of numerical techniques that are not only accurate and cost-effective but also guarantee physical consistency. These methods should adeptly capture the interface, seamlessly integrating it with the momentum conservation equation. Unfortunately, developing such approaches proves to be complex due to the inherent challenges associated with multiphase flows. These difficulties encompass a range of issues, including but not restricted to Bestion: (1) Upholding the conservation of mass, momentum, and kinetic energy, (2) Modeling variations in properties across interfaces, particularly significant shifts in density and viscosity, (3) Handling intricate topologies and the differentiation of scales, (4) Ensuring stability in simulating multiphase flows, and (5) Precisely incorporating surface tension forces while maintaining accuracy [3-5].

Figure 1 presents an overview of various categories within two-phase modeling, emphasizing prominent methods that currently garner significant attention in the community. While two-fluid models prove effective for addressing simple problems, they often fall short when handling realistic scenarios [2]. In-depth analyses of two-phase flow models are extensively covered in the works of Ishii and Hibiki and Prosperetti & Tryggvason [2,6]. For readers intrigued by Marker-and-Cell (MAC) techniques, comprehensive discussions can be found in the works of McKee et al. [7]. Similarly, those interested in Front-Tracking (FT) methods can refer to the comprehensive discussions provided by Tryggvason et al. [8,9]. Additionally, the realm of diffuse-interface approaches encompasses Phase Field (PF), Constrained Interpolation Profile (CIP), and Conservative Level-Set (CLS) methods. In contrast, sharp-interface approaches involve different classes of Interface-Capturing and Interface-Tracking methods. It is important to highlight certain abbreviations in Figure 1, such as SPH, LBM, LS, and CLSVOF, which represent the Smoothed-Particle Hydrodynamics, Lattice Boltzmann Method, Level-Set and Coupled Level-Set and Volume-of-Fluid [10-13], respectively.

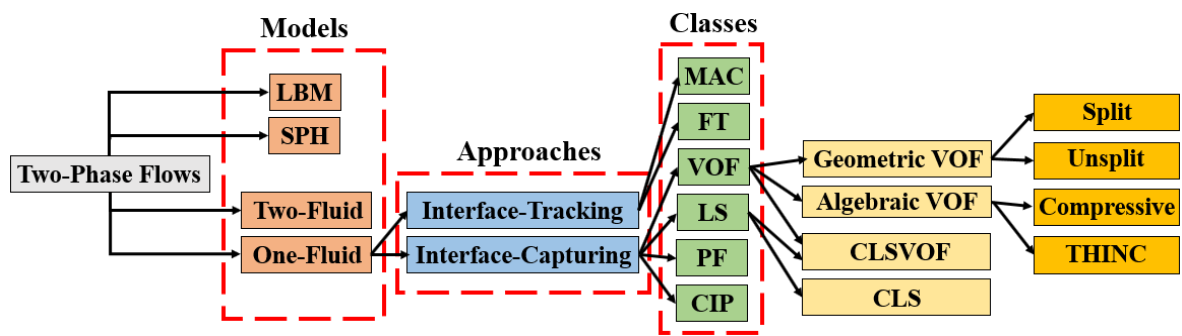

**Figure 1:** Categorization of Numerical Techniques for Handling Two-Phase Flows. The Definitions of Abbreviations can be Found in the Main Text

This study primarily centers on one-fluid models, methodologies for capturing interfaces, and Volume-of-Fluid (VOF) techniques [4]. In the VOF approach, the interface delineation occurs implicitly through the allocation of volume fractions to represent the presence of a specific fluid within computational cells. In this context, advection is accomplished by redistributing the fluid's content among neighboring cells, facilitated by its movement across the faces of the computational mesh. Since the introduction of the original VOF techniques recorded in the literature numerous varied VOF approaches have been conceptualized and created [14]. According to Figure 1, there exist two overarching classifications within this method: geometric methodologies involve a detailed reconstruction of the interface using data related to volume fractions [12,13]. On the other hand, algebraic methodologies avoid explicit reconstruction efforts. Algebraic schemes, renowned for their relative simplicity of implementation, enhanced efficiency, and adaptability to unstructured meshes, operate without the constraints of structured meshes. Nevertheless, their foundation relies on heuristic considerations, rendering them less accurate in comparison to their geometric counterparts [15]. Conversely, geometric VOF schemes necessitate intricate geometric operations, resulting in a more cumbersome implementation process and slower execution [5]. It is worth noting that the pursuit of geometric VOF methods tailored for unstructured meshes constitutes a dynamic and ongoing area of research [16-22]. Illustrated in Figure 1, algebraic VOF methods are categorized into two groups: 1) THINC and 2) compressive.

THINC schemes, as proposed by Xiao et al. encompass a collection of recently developed techniques within algebraic VOF methods [23]. These methods assume a hyperbolic-tangent profile for the volume fraction within the computational cell containing the interface and obtain the fluxes algebraically. As asserted by Xie et al. these approaches exhibit accuracy levels similar to those of geometric schemes but with a lower computational cost [21]. As a result, THINC schemes are increasingly garnering attention in current research. Additionally, unlike compressive schemes, these methods eliminate necessity for artificial compression and are not dependent on the Courant number at the local cell level [24]. The second group is also known as Interface Compression (IC) methods, which are high-resolution schemes. These formulations integrate an artificial compression factor into the advection equation for volume fraction, leading to an unconventional diffusion coefficient with a negative value. This term functions by constricting the volume fraction distribution perpendicular to the fluid interface, thereby mitigating interface dispersal resulting from numerical diffusion. Consequently, it supports the limitations and preservation of the proportion of phases [25]. The compressive impact varies based on the mesh resolution as well as intrinsic IC coefficient, which represents a parameter linked to the artificial compression term. Various investigations have been carried out to examine the grid resolution and the coefficient of the interaction, as outlined in prior studies [15,26-28].

Deshpande et al. identified the occurrence of parasitic currents induced by the implementation of the IC technique in flows primarily governed by the effects of interfacial tension [15,29]. Parasitic flows manifest as resilient abstract vortices in close proximity to the interface, arising autonomously without external influences due to inaccuracies in calculating interface curvature or disparities between surface tension and pressure gradient forces. Typically, these issues manifest in simulations of static bubbles and droplets within fluids characterized by high-density ratios [30]. Hoang et al. investigated how the IC coefficient influences various factors such as highest velocity attained, the thickness of the VOF interface, and the occurrence of parasitic currents [31]. They convincingly demonstrated that the optimal condition for preventing parasitic flows along with numerical diffusion corresponds to an IC coefficient of 1. Furthermore, their research emphasized the importance of cell dimensions in determining the optimal conditions for IC coefficient.

In this investigation, our aim is to utilize a precise algebraic approach for resolving the interface advection equation within the context of two-phase flow. This method is commonly known as the Multidimensional Universal Limiter for Explicit Solution (MULES) algorithm [15]. Our approach involves implementing this method through Python coding, as it is currently accessible within the OpenFOAM software. While this open-source software provides users with convenient utilization of this method, it has yet to be made available as an in-house code. The rest of the article is organized as follows: Firstly, the problem statement and modeling approach are introduced in sections 2.1 to 2.3. Subsequently, section 2.4 provides details on the numerical solution procedure, including the discretization of equations. The verification follows in section 3, where results from various schemes, such as first-order upwind second-order central van Leer flux limiter, and geometric VOF, are compared.

Section 4.1 explores the impact of changing the IC coefficient on the method's accuracy [9,32-34]. In section 4.2, the value of surface tension is determined using the results of interface capturing with two methods, MULES and van Leer flux limiter (as verification), and its accuracy is assessed. Sections 3 and 4 include criteria such as Interface Advection Error (IAE), Mass Conservation Error (MCE), Elapsed Time (ET), Dimensionless Pressure Difference (DPD), and Capillary number (Ca). The article concludes with a summary of major findings in section 5.

## 2. Physical and Mathematical Modelling
### 2.1 Problem Statement
In Figure 2, a bubble with a diameter $D_0$=0.3 $m$ is positioned at coordinates ($x_c$ = 0.5 m, $y_c$=0.75 m) within a square domain measuring 1 m in length ($l$=1 $m$). A two-dimensional velocity field, featuring a periodicity of $T$ = 2 s, is applied to induce a two-dimensional (2D) vortex. Within this dynamic velocity field, the droplet undergoes a reciprocating rotational motion throughout a 2-second interval [35]. The volume fraction of the droplet is denoted as α and is defined as 1, while its immediate surroundings exhibit a volume fraction of 0. Also, the interface of the droplet is precisely characterized by a volume fraction of 0.5.

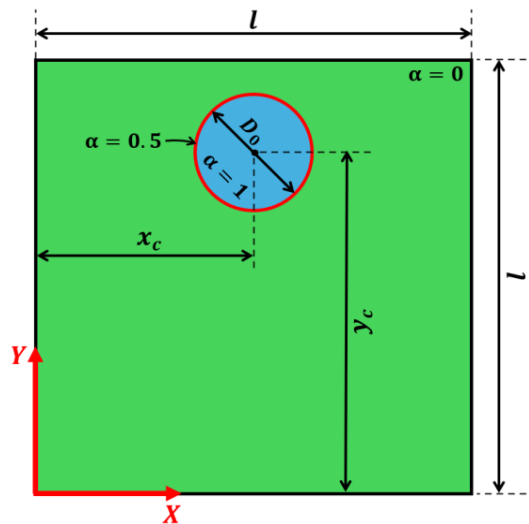

**Figure 2:** The Computational Domain; Regions Colored in Green and Blue are the Surroundings and Droplet, Respectively.

### 2.2 Governing Equations
In multiphase flow modeling using the VOF method as the one-fluid formulation [9], the volume fraction advection equation is solved according to equation (1), along with the continuity and momentum equations [12]. Typically, this equation is solved for the secondary phase. Since the sum of the volume fractions of all phases is equal to 1, the volume fraction of the primary phase fluid can be obtained by a simple subtraction operation after solving the equation for the volume fraction of the secondary phase, as shown in equation (2).

$$\frac{\partial \alpha_q}{\partial t} + \nabla . (U \alpha_q) = 0 \qquad (1)$$

$$\alpha_p = 1 - \alpha_q \qquad (2)$$

where α is the volume fraction, and the subscripts $p$ and $q$ refer to the primary (surroundings) and secondary (droplet) phases, respectively. Also, tensorial quantities are indicated by bold symbols. For example, $U$ is the velocity vector field created by 2D vortex.

The accurate calculation of the volume fraction is crucial because it determines the curvature of the interface, surface tension, and pressure gradients. However, in multiphase flow problems involving phases with high density ratios, even small errors in the calculation and distribution of the volume fraction can

J Math Techniques Comput Math, 2024                                                                                                          Volume 3 | Issue 10 | 4

cause significant and severe changes in the effective properties. Additionally, since the phase interface is confined to only a few computational cells, the simulation's dependency on grid size increases within this range when using the VOF method. Rusche [36] introduced a convective term to equation (1), which originates from using the velocity field derived from the weighted average of the velocities of all phases. This weighted average is defined according to equation (3) under the assumption that the influence of each phase's speed on interface changes is directly related to the volume fraction of that phase [37].

$$\boldsymbol{U} = \alpha_q \boldsymbol{U}_q + (1 - \alpha_q)\boldsymbol{U}_p \tag{3}$$

To integrate equations (1) and (3), the relative velocity, $\boldsymbol{U}_r$, between the two phases is defined as follows [38]:

$$\boldsymbol{U}_r = \boldsymbol{U}_p - \boldsymbol{U}_q \tag{4}$$

Therefore, by combining equations (1), (3), and (4), the advection equation that governs the volume fraction of the fluid used in the MULES algorithm can be expressed as follows [37]:

$$\frac{\partial \alpha_q}{\partial t} + \nabla \cdot (\boldsymbol{U}\alpha_q) + \nabla \cdot [\boldsymbol{U}_r \alpha_q (1 - \alpha_q)] = 0 \tag{5}$$

The third term in this equation does not have a specific physical meaning but is used to compress the phase interface during the numerical solution of the volume fraction conservation equation. Notably, this term equals zero at the lower and upper limits of the fluid volume fraction, i.e., 0 and 1, and only operates at the interface. Consequently, it more accurately tracks the interface and reduces numerical errors resulting from numerical diffusion. Now that the problem of the numerical solution of the volume fraction equation has been solved, the surface tension force can be calculated based on the volume fraction gradient. This force arises due to the excess pressure gradient at the interface between phases [39]. Brackbill et al. proposed using the Continuum Surface Force (CSF) method to model it [40]. In this method, the surface tension force per unit volume of the fluid element is calculated using the following equation:

$$\boldsymbol{F}_\sigma = \sigma \kappa \nabla \alpha_q \tag{6}$$

Where $\sigma$ and $\kappa$ represent the surface tension coefficient and the mean radius of interface curvature, respectively. The mean radius of interface curvature is defined as follows:

$$\kappa = -\nabla \cdot \left(\frac{\nabla \alpha_q}{|\nabla \alpha_q|}\right) \tag{7}$$

It should be noted that the given flow velocity components in relation to the stream function, $\psi$, must be [41]:

$$u = \frac{\partial \psi}{\partial y} \tag{8}$$

$$v = -\frac{\partial \psi}{\partial x} \tag{9}$$

Where $\psi$ can be defined as follows [35,42]:

$$\psi(x, y, t) = \frac{1}{\pi} \cos\left(\frac{\pi t}{T}\right) \sin^2\left(\frac{\pi x}{l}\right) \sin^2\left(\frac{\pi y}{l}\right) \tag{10}$$

Therefore, there is no need to solve the continuity and momentum equations to obtain velocity field.

### 2.3 Dimensionless Parameters
The objective parameters described in the previous sections can be alternatively expressed as [5,17,43,44]:

$$\text{IAE} = \sum_{i,j}(\alpha_{i,j}^T - \alpha_{i,j}^0)dxdy \qquad (11)$$

$$\text{MCE} = \frac{|V^T - V^0|}{V^0} = \frac{|\sum_{i,j}\alpha_{i,j}^T dxdy - \sum_{i,j}\alpha_{i,j}^0 dxdy|}{\sum_{i,j}\alpha_{i,j}^0 dxdy} \qquad (12)$$

$$\text{ET} = t^T - t^0 \qquad (13)$$

$$\text{DPD} = \frac{D_0 \Delta p}{2\sigma} \qquad (14)$$

$$\text{Ca} = \frac{\mu_q U_{max}}{\sigma} \qquad (15)$$

In these parameters, the subscripts, $j$ and the superscripts of 0 and $T$ represent the 2D computational cell in the domain, the initial state, and the final state, respectively. Additionally, $V$ denotes the volume of the droplet, $\mu$ represents the dynamic viscosity, and $U_{max}$ indicates the maximum velocity of the parasitic current. In this study, the density ratio, $\frac{\rho_q}{\rho_p}$, the dynamic viscosity ratio, $\frac{\mu_q}{\mu_p}$, and the surface tension coefficient are considered to be 20, 1, and 0.1 $N/m$, respectively. Based on this information, three governing (design) dimensionless parameters can be defined as follows [5]:

$$\text{Ar} = \frac{g\rho_q \Delta\rho D_0^3}{\mu_q^2} = 1.03 \times 10^4 \qquad (16)$$

$$\text{Eo} = \frac{\Delta\rho g D_0^2}{\sigma} = 171 \qquad (17)$$

$$\text{Mo} = \frac{\text{Eo}^3}{\text{Ar}^4} = 4.44 \times 10^{-10} \qquad (18)$$

Where Ar, Eo, Mo, and are Archimedes, Eotvos, and Morton numbers, respectively. Furthermore, equation (14) is derived using the analytical solution for pressure difference in a two-dimensional domain. Therefore, the value of the DPD should be equal to 1 in the ideal mode, making it a reliable criterion for evaluating different schemes in section 4.2 [41].

## 2.4 Numerical Solution
The schematic of computational grid is shown in Figure 3. The quantities of pressure and volume fraction are stored at the centers of the cells, while the velocity components are stored at their faces. Interpolation is used to obtain the desired quantity whenever necessary.

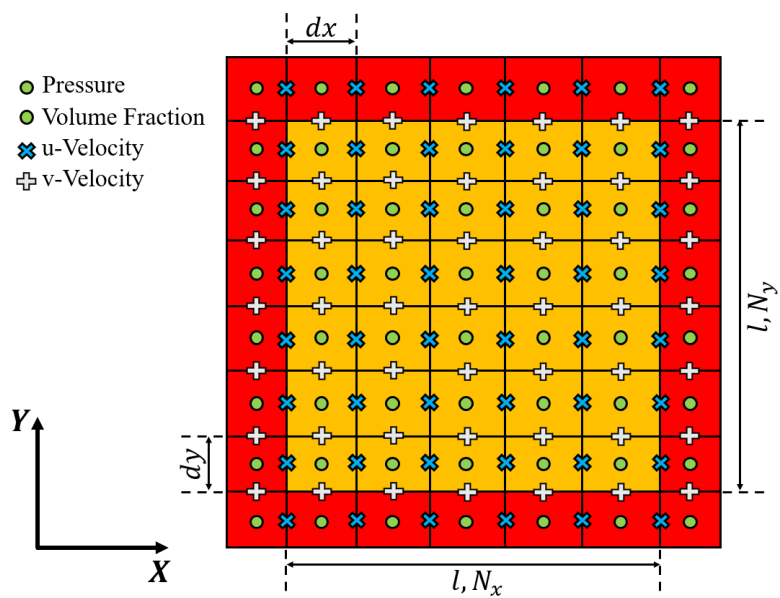

**Figure 3:** The Computational Grid; Cells Colored in Orange and Red are the Main and Ghost Cells, Respectively

Now it is time to perform the volume integration on both sides of equation (5), which is as follows [9]:

$$\int_V \frac{\partial \alpha_q}{\partial t} dV + \int_V \nabla \cdot [U\alpha_q + U_r \alpha_q(1-\alpha_q)]dV = 0 \qquad (19)$$

By applying the divergence theorem on the advection term and using the explicit Euler method for the unsteady term, the discretized form can be obtained as follows [15]:

$$\alpha_{q_{i,j}}^{n+1} = \alpha_{q_{i,j}}^n - \frac{\Delta t}{V_{i,j}} \sum_f F_f^n \qquad (20)$$

Here, the subscript of $f$ and the superscript of $n$ represent the cell face and the current time step, respectively. Furthermore, $\Delta t$, $V_{i,j}$, and $F_f^n$ are time step size, volume of the computational cell, and advective flux on the cell face. The advection term is expressed as a summation over the cell faces. The cell volume and advective flux can be obtained as follows:

$$V_{i,j} = (dxdydz)_{i,j} \xrightarrow{2D} V_{i,j} = (dxdy)_{i,j} \qquad (21)$$

$$F_f^n = F_u^n + \lambda_m F_c^n \qquad (22)$$

Where $F_u^n$, $F_c^n$, and $\lambda_m$ denote the first-order upwind flux, corrected flux, and MULES limiter, respectively. The fluxes $F_u^n$ and $F_c^n$ are expressed by [15]:

$$F_u^n = \left(\phi_f \alpha_{q_{f,U}}\right)^n \qquad (23)$$

$$F_c^n = \left(\phi_f \alpha_{q_f}\right)^n + \left[\phi_{rf} \alpha_{q_{rf}}(1-\alpha_{q_{rf}})\right]^n - F_u^n \qquad (24)$$

Where the volume flux, $\phi_f$ and corrected volume flux, $\phi_{rf}$, are assigned by:

$$\phi_f = U_f \cdot S_f = |U_f| n_f \cdot S_f \qquad (25)$$

$$\phi_{rf} = U_{rf} \cdot S_f = \min_{f \in V_{i,j}} \left[ IC \frac{|\phi_f|}{|S_f|}, \max_{f \in Domain}\left(\frac{|\phi_f|}{|S_f|}\right)\right] n_f \cdot S_f \qquad (26)$$

It is important to note that in equation (26), the maximum operation is applied across all faces in the domain, whereas the minimum operation is performed locally within each cell. In addition, in equations (25) and (26), $U_f$, $n_f$, and $S_f$ represent the face velocity, the unit normal vector to the interface, and the surface area vector of the cell face, respectively. These quantities can be determined by:

$$U_f = \frac{U_P + U_N}{2} \qquad (27)$$

$$n_f = \frac{(\nabla \alpha_q)_f}{|(\nabla \alpha_q)_f| + \frac{10^{-8}}{\sqrt[3]{V_{i,j}}}} \qquad (28)$$

$$S_f = \begin{cases} dy\, \mathbf{i} & ; \text{vectival face} \\ dx\, \mathbf{j} & ; \text{horizontal face} \end{cases} \qquad (29)$$

In equation (26), the variable IC is typically treated as a constant value ranging from 0 to 2 in order to limit interface smearing. Furthermore, $\alpha_{q_{f,U}}$ in equation (23), along with $\alpha_{q_f}$ and $\alpha_{q_{rf}}$ in equation (24), can be calculated using the relations below [5,15].

$$\alpha_{q_{f,U}} = \begin{cases} \alpha_{q_P} & ; \text{if } \phi_f \geq 0 \\ \alpha_{q_N} & ; \text{if } \phi_f < 0 \end{cases} \quad (30)$$

$$\alpha_{q_f} = \alpha_{q_P} + \frac{\alpha_{q_N} - \alpha_{q_P}}{2}[1 - \xi(\phi_f)(1 - \omega_f)] \quad (31)$$

$$\alpha_{q_{rf}} = \alpha_{q_P} + \frac{\alpha_{q_N} - \alpha_{q_P}}{2}[1 - \xi(\phi_f)(1 - \omega_{rf})] \quad (32)$$

Where $\xi(\phi_f)$, $\omega_f$, and $\omega_{rf}$ represent the step function, van Leer's flux limiter function and the flux limiter of the IC scheme, respectively [33]. Their definitions are as follows:

$$\xi(\phi_f) = \begin{cases} 1 & ; \text{if } \phi_f \geq 0 \\ -1 & ; \text{if } \phi_f < 0 \end{cases} \quad (33)$$

$$\omega_f = \omega(r_f) = \left(\frac{r+|r|}{1+|r|}\right)_{r=r_f} \quad (34)$$

$$\omega_{rf} = \min\left[\max\left[1 - \max\left[\left(1 - \left(4\alpha_{q_P}(1 - \alpha_{q_P})\right)\right)^2, \left(1 - \left(4\alpha_{q_N}(1 - \alpha_{q_N})\right)\right)^2\right], 0\right], 1\right] \quad (35)$$

In equation (34), $r_f$ denotes the ratio of successive gradients within the computational domain (refer to Figure 3) and can be defined as follows:

$$r_f = \frac{\alpha_{q_U} - \alpha_{q_{UU}}}{\alpha_{q_D} - \alpha_{q_U}} \quad (36)$$

In equations above, the subscripts *P, N, U, UU,* and *D* correspond to the current cell, neighbor cell, first upwind cell, second upwind cell, and first downwind cell, respectively. To enhance understanding, a row of these cells is illustrated in Figure 4.

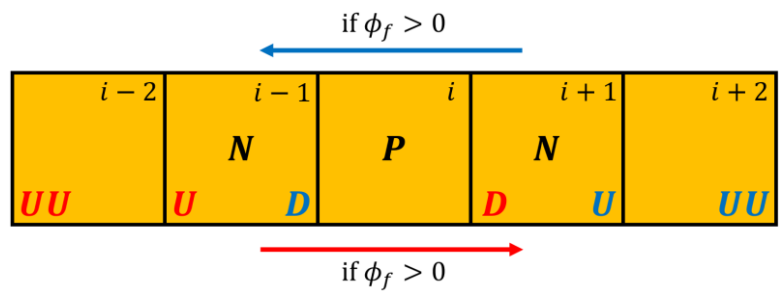

**Figure 4:** The Arrangement of Important Computing Cells Including Current (*P*), Neighbor (*N*), First Upwind (*U*), Second Upwind (*UU*), and First Downwind (*D*) Cells

Returning to equation (22), $\lambda_m$ is incorporated in the MULES algorithm and is equal to one in the transition region while being zero elsewhere. To compute this quantity, it is necessary to determine the local extrema of the volume fraction, as described below (refer to Figure 5) [45,46]:

$$\alpha_q^{min} = \min\left(\alpha_{q_P}^n, \alpha_{q_N}^n\right) \xrightarrow{2D} \alpha_{q_{i,j}}^{min} = \min\left(\alpha_{q_{i,j}}^n, \alpha_{q_{i+1,j}}^n, \alpha_{q_{i-1,j}}^n, \alpha_{q_{i,j+1}}^n, \alpha_{q_{i,j-1}}^n\right) \quad (37)$$

$$\alpha_q^{max} = \max\left(\alpha_{q_P}^n, \alpha_{q_N}^n\right) \xrightarrow{2D} \alpha_{q_{i,j}}^{max} = \max\left(\alpha_{q_{i,j}}^n, \alpha_{q_{i+1,j}}^n, \alpha_{q_{i-1,j}}^n, \alpha_{q_{i,j+1}}^n, \alpha_{q_{i,j-1}}^n\right) \quad (38)$$

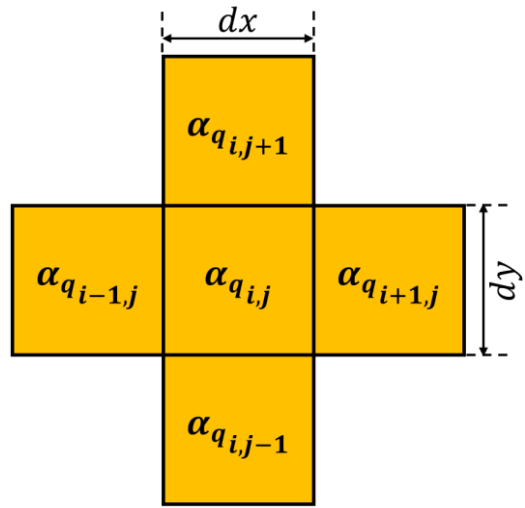

**Figure 5:** The Schematic for Calculating the Local Extrema of the Volume Fraction; Current and Neighbor Cells

To adhere to the lower and upper limits of the volume fraction and to minimize numerical error, it is essential to adjust equations (37) and (38) as follows:

$$\alpha_q^{min,c} = \max(\alpha_q^{min}, 0) \tag{39}$$

$$\alpha_q^{max,c} = \min(\alpha_q^{max}, 1) \tag{40}$$

Subsequently, as shown in Figure 6, the magnitudes of the inflows and outflows of anti-diffusive fluxes from each face of the cell, denoted as $A_f^{\mp}$, along with their cumulative sum for each cell, $P^{\pm}$, must be calculated according to the following relations. In $A_f^{\mp}$, the negative and positive superscripts indicate inflows and outflows, respectively [46,47].

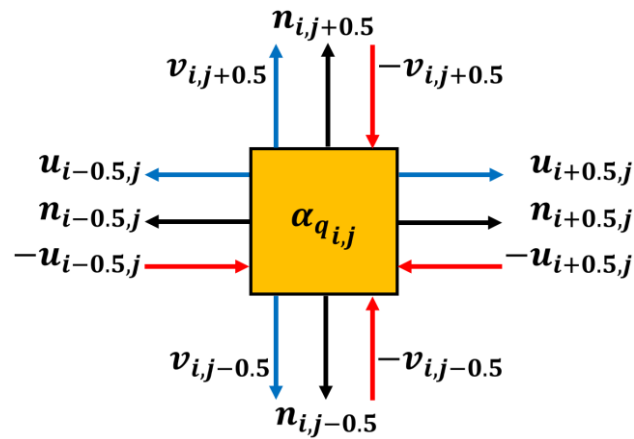

**Figure 6:** The Schematic for Calculating the Inflows and Outflows of Anti-Diffusive Fluxes from each Cell Face, Along with their Cumulative Sum for Each Cell

$$A_f^{\mp} = (F_h^n - F_u^n)^{\mp} \xrightarrow{2D} \begin{cases} A_{i,j+0.5}^{+} = v_{i,j+0.5}^n \left[\frac{\alpha_{q_{i,j+1}}^n + \alpha_{q_{i,j}}^n}{2} - \alpha_{q_{i,j}}^n\right] dx & ; \text{if } v_{i,j+0.5}^n \geq 0 \\ A_{i,j+0.5}^{-} = v_{i,j+0.5}^n \left[\frac{\alpha_{q_{i,j+1}}^n + \alpha_{q_{i,j}}^n}{2} - \alpha_{q_{i,j+1}}^n\right] dx & ; \text{if } v_{i,j+0.5}^n < 0 \\ A_{i,j-0.5}^{+} = v_{i,j-0.5}^n \left[\frac{\alpha_{q_{i,j-1}}^n + \alpha_{q_{i,j}}^n}{2} - \alpha_{q_{i,j}}^n\right] dx & ; \text{if } v_{i,j-0.5}^n < 0 \\ A_{i,j-0.5}^{-} = v_{i,j-0.5}^n \left[\frac{\alpha_{q_{i,j-1}}^n + \alpha_{q_{i,j}}^n}{2} - \alpha_{q_{i,j-1}}^n\right] dx & ; \text{if } v_{i,j-0.5}^n \geq 0 \\ A_{i+0.5,j}^{+} = u_{i+0.5,j}^n \left[\frac{\alpha_{q_{i+1,j}}^n + \alpha_{q_{i,j}}^n}{2} - \alpha_{q_{i,j}}^n\right] dy & ; \text{if } u_{i+0.5,j}^n \geq 0 \\ A_{i+0.5,j}^{-} = u_{i+0.5,j}^n \left[\frac{\alpha_{q_{i+1,j}}^n + \alpha_{q_{i,j}}^n}{2} - \alpha_{q_{i+1,j}}^n\right] dy & ; \text{if } u_{i+0.5,j}^n < 0 \\ A_{i-0.5,j}^{+} = u_{i-0.5,j}^n \left[\frac{\alpha_{q_{i-1,j}}^n + \alpha_{q_{i,j}}^n}{2} - \alpha_{q_{i,j}}^n\right] dy & ; \text{if } u_{i-0.5,j}^n < 0 \\ A_{i-0.5,j}^{-} = u_{i-0.5,j}^n \left[\frac{\alpha_{q_{i-1,j}}^n + \alpha_{q_{i,j}}^n}{2} - \alpha_{q_{i-1,j}}^n\right] dy & ; \text{if } u_{i-0.5,j}^n \geq 0 \end{cases} \quad (41)$$

$$P^{\pm} = \mp \sum_f A_f^{\mp} \xrightarrow{2D} \begin{cases} P_{i,j}^{+} = -[A_{i,j+0.5}^{-} + A_{i,j-0.5}^{-} + A_{i+0.5,j}^{-} + A_{i-0.5,j}^{-}] \\ P_{i,j}^{-} = [A_{i,j+0.5}^{+} + A_{i,j-0.5}^{+} + A_{i+0.5,j}^{+} + A_{i-0.5,j}^{+}] \end{cases} \quad (42)$$

Where $F_h^n$ represents the central-differencing flux. Next, it is essential to calculate the total transported flux, which arises from the differences in the volume fraction of each cell in relation to the local extrema of the volume fraction, along with the sum of the first-order upwind flux across all faces, as follows [5,46]:

$$Q_{i,j}^{+} = \frac{V_{i,j}}{\Delta t}\left(\alpha_q^{max,c} - \alpha_{q_{i,j}}^n\right) + \sum_f F_{u,f}^n \quad (43)$$

$$Q_{i,j}^{-} = \frac{V_{i,j}}{\Delta t}\left(\alpha_{q_{i,j}}^n - \alpha_q^{min,c}\right) - \sum_f F_{u,f}^n \quad (44)$$

By introducing an internal loop denoted by $v$, the MULES limiter can be determined for all faces and the center of the cells as follows [5, 46]:

$$\lambda_{m,i,j}^{\mp,v+1,n} = \max\left[\min\left(\frac{\pm \sum_f \lambda_{m,f}^{v,n} A_f^{\pm} + Q_{i,j}^{\pm}}{P_{i,j}^{\pm}}, 1\right), 0\right] \quad (45)$$

$$\lambda_{m,f}^{v+1,n} = \begin{cases} \min(\lambda_{m,P}^{+,v+1,n}, \lambda_{m,N}^{-,v+1,n}) & ; \text{if } A_f \geq 0 \\ \min(\lambda_{m,P}^{-,v+1,n}, \lambda_{m,N}^{+,v+1,n}) & ; \text{if } A_f < 0 \end{cases} \quad (46)$$

In this iterative process, the initial value of MULES limiter for all faces is set to 1 ($\lambda_{m,f}^{v=1,n}=1$). This loop continues until $v$ reaches either 2 or 3. After discretizing equation (5) and implementing the MULES algorithm, a numerical solution must be developed for equations (6) and (7). In this study, the delta function approximation is employed to model the standard Continuum Surface Force (CSF) [4,9]. As illustrated in Figure 7, the mean radius of curvature at the interface is stored at the cell center, while the unit normal vector to the interface is stored at the cell faces. Additionally, according to equation (28), the interface normal vector is directly related to the gradient of the volume fraction. Therefore, it is also anticipated that this gradient, along with the surface tension, will be stored at the cell faces. The approximated form of the surface tension is given by:

$$\boldsymbol{F_\sigma} = \boldsymbol{f_\sigma}\delta_s = -\sigma\kappa\boldsymbol{\nabla}\alpha_q \sim -\sigma\kappa\boldsymbol{\nabla_h}\alpha_q \quad (47)$$

Here $\nabla_h$ denotes the discrete gradient operator. For clarity, equation (47) is discretized for the right face at $(i + 0.5, j)$ as follows [5]:

$$F_{\sigma,x,i+0.5,j} = -\sigma \kappa_{i+0.5,j} \left(\frac{\partial \alpha_q}{\partial x}\right)_{i+0.5,j} = -\sigma \left(\frac{\kappa_{i,j}^n + \kappa_{i+1,j}^n}{2}\right)\left(\frac{\alpha_{q_{i+1,j}}^n - \alpha_{q_{i,j}}^n}{dx}\right) \quad (48)$$

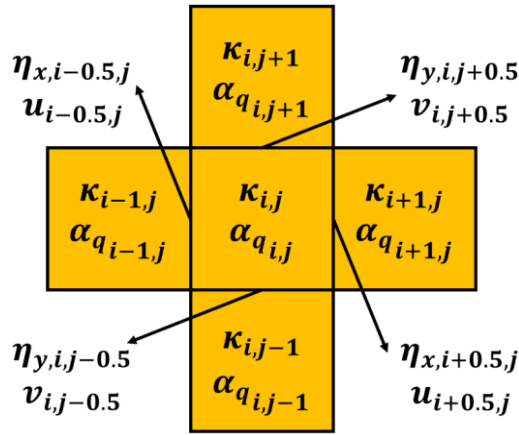

**Figure 7:** The Schematic for Calculating the Mean Radius of Curvature at the Interface and the Associated Surface Tension

The mean radius of curvature at the interface, denoted as $\kappa_{i,j}^n$, can be determined using the following equation:

$$\kappa_{i,j}^n = -\left[\frac{1}{dx}\left(\eta_{x,i+0.5,j}^n - \eta_{x,i-0.5,j}^n\right) + \frac{1}{dy}\left(\eta_{y,i,j+0.5}^n - \eta_{y,i,j-0.5}^n\right)\right] \quad (49)$$

Here, $\eta_x^n$ represents the discretized form of the unit normal vector to the interface, which is expressed as follows:

$$\eta_{x,i+0.5,j}^n = \frac{m_{x,i+0.5,j}^n}{\sqrt{\left(m_{x,i+0.5,j}^n\right)^2 + \left(m_{y,i+0.5,j}^n\right)^2 + \frac{10^{-8}}{\sqrt[3]{V_{i,j}}}}} \quad (50)$$

In equation (50), $m_x^n$ and $m_y^n$ represent the components of the volume fraction gradient, defined as follows:

$$m_{y,i+0.5,j}^n = 0.25\left[m_{y,i,j+0.5}^n + m_{y,i+1,j+0.5}^n + m_{y,i,j-0.5}^n + m_{y,i+1,j-0.5}^n\right] \quad (51)$$

$$m_{y,i,j+0.5}^n = \frac{\alpha_{q_{i,j+1}}^n - \alpha_{q_{i,j}}^n}{dy} \quad (52)$$

$$m_{x,i+0.5,j}^n = \frac{\alpha_{q_{i+1,j}}^n - \alpha_{q_{i,j}}^n}{dx} \quad (53)$$

It is important to note that these calculations are straightforward for the other faces as well. A key aspect of calculating surface tension is its dependence on volume fraction. Smoothing this quantity through filtering may enhance accuracy. This filtering method can also effectively impact critical parameters in this study [5,9]. As shown in Table 1, this process is implemented in multiple steps.

| Tag | Name of Step | Smoothing the volume fraction | |
|---|---|---|---|
| | | Equation | |
| C1 | First Step of Filtering | $\widetilde{\alpha}_{q_{i,j}}^{n} = \frac{1}{2}\alpha_{q_{i,j}}^{n} + \frac{1}{8}\left(\alpha_{q_{i+1,j}}^{n} + \alpha_{q_{i,j+1}}^{n} + \alpha_{q_{i-1,j}}^{n} + \alpha_{q_{i,j-1}}^{n}\right)$ | (54) |
| C2 | Second Step of Filtering | $\widetilde{\widetilde{\alpha}}_{q_{i,j}}^{n} = \frac{1}{2}\widetilde{\alpha}_{q_{i,j}}^{n} + \frac{1}{8}\left(\widetilde{\alpha}_{q_{i+1,j}}^{n} + \widetilde{\alpha}_{q_{i,j+1}}^{n} + \widetilde{\alpha}_{q_{i-1,j}}^{n} + \widetilde{\alpha}_{q_{i,j-1}}^{n}\right)$ | (55) |
| C3 | Third Step of Filtering | $\widetilde{\widetilde{\widetilde{\alpha}}}_{q_{i,j}}^{n} = \frac{1}{2}\widetilde{\widetilde{\alpha}}_{q_{i,j}}^{n} + \frac{1}{8}\left(\widetilde{\widetilde{\alpha}}_{q_{i+1,j}}^{n} + \widetilde{\widetilde{\alpha}}_{q_{i,j+1}}^{n} + \widetilde{\widetilde{\alpha}}_{q_{i-1,j}}^{n} + \widetilde{\widetilde{\alpha}}_{q_{i,j-1}}^{n}\right)$ | (56) |
| C4 | Forth Step of Filtering | $\widetilde{\widetilde{\widetilde{\widetilde{\alpha}}}}_{q_{i,j}}^{n} = \frac{1}{2}\widetilde{\widetilde{\widetilde{\alpha}}}_{q_{i,j}}^{n} + \frac{1}{8}\left(\widetilde{\widetilde{\widetilde{\alpha}}}_{q_{i+1,j}}^{n} + \widetilde{\widetilde{\widetilde{\alpha}}}_{q_{i,j+1}}^{n} + \widetilde{\widetilde{\widetilde{\alpha}}}_{q_{i-1,j}}^{n} + \widetilde{\widetilde{\widetilde{\alpha}}}_{q_{i,j-1}}^{n}\right)$ | (57) |

**Table 1: The Process of Smoothing the Volume Fraction, Along with its Corresponding Steps and Equations**

## 2.5 Verification

In this section, the verification of the MULES algorithm using a spherical droplet in a 2D vortex ($N_x = N_y = 32$ and $\Delta t = 0.005$s), as detailed in section 2.1, is examined. The results, including IAE, MCE, and ET, are compared with those obtained from the first-order upwind, second-order central, van Leer flux limiter, and geometric VOF methods. First, it is important to note that if the MULES limiter, $\lambda_m$, in Equation (22) is set to 0, the results of the MULES solver should exactly match those of the first-order upwind solver. Table 2 presents three parameters including interface advection error, mass conservation error, and elapsed time (in seconds) for both the MULES and first-order upwind methods.

| Parameter | MULES $IC = 0.1$ | MULES $IC = 0.5$ | First-Order Upwind |
|---|---|---|---|
| IAE (%) | 9.61 | 9.61 | 9.61 |
| MCE (%) | 0 | 0 | 0 |
| ET (s) | 153.59 | 150.08 | 108.76 |

**Table 2: The Comparison Between the two Interface Capturing Strategies, MULES and First-Order Upwind, Based on Three Criteria: IAE, MCE, and ET**

As observed in Table 2, for two different values of IC in the MULES scheme, the results for the specified parameters are consistent with those of the first-order upwind method, confirming the accuracy of the MULES solver. The slight difference in elapsed time (approximately 40 seconds) between the two methods is due to the more complex coding of the MULES solver and its use of different functions, which require more time to achieve the same results as the first-order upwind method. According to Amani [5], the MULES method has a lower computational cost than the geometric VOF method, but the geometric VOF method offers higher accuracy. It is also important to ensure that the MULES method achieves good convergence and stability, and performs more accurately than other methods such as first-order upwind, second-order central, and van Leer flux limiter. To demonstrate this, the results for these methods are provided in Table 3. Additionally, Figure 8a-e displays the volume fraction contours for various schemes at three selected times $(0, T/2, T)$.

| Parameter | MULES $IC = 0.1$ | MULES $IC = 0.5$ | First-Order Upwind | Second-Order Central | van Leer Flux Limiter | Geometric VOF [5] |
|---|---|---|---|---|---|---|
| IAE (%) | 4.63 | 4.22 | 9.61 | 4.20 | 4.82 | 0.63 |
| MCE (%) | 0.82 | 5.64 | 0 | 10.68 | 0 | 0 |
| ET (s) | 146.51 | 149.82 | 108.76 | 109.43 | 111.61 | - |

**Table 3: The Comparison between the Different Interface Capturing Schemes, Based on Three Criteria: IAE, MCE, and ET**

As shown in Table 3, the computational cost of the MULES method is higher than that of the other methods (approximately 40 seconds). This is expected due to the more complex calculations and formulations involved in MULES. While the accuracy of the MULES method is lower than that of the geometric method, it is on par with, or even better than, the other methods. This is confirmed by comparing the percentages of IAE and MCE. For example, the IAE in the first-order upwind method is nearly twice that of the MULES, second-order central, and van Leer flux limiter methods. Consequently, this method is likely to experience significant numerical diffusion, leading to inaccurate interface tracking (see Figure 8c), whereas the other methods achieve more consistent interface tracking. Additionally, the MCE for the first-order upwind, van Leer flux limiter, and geometric VOF methods is zero, ensuring high stability and preventing interface noise due to mass differences. In contrast, the MCE for the second-order central difference method is 10.68%, which is 13 times and 1.89 times higher than the MULES method with two different IC values of 0.1 and 0.5, respectively. This higher error indicates that the second-order central difference method is prone to fluctuations at the phase interface, resulting in lower stability (see Figure 8d). Therefore, both the MULES (see Figure 8a-b) (Multimedia available online) and van Leer flux limiter (see Figure 8e) schemes are stable numerical approaches that accurately track the phase interface.

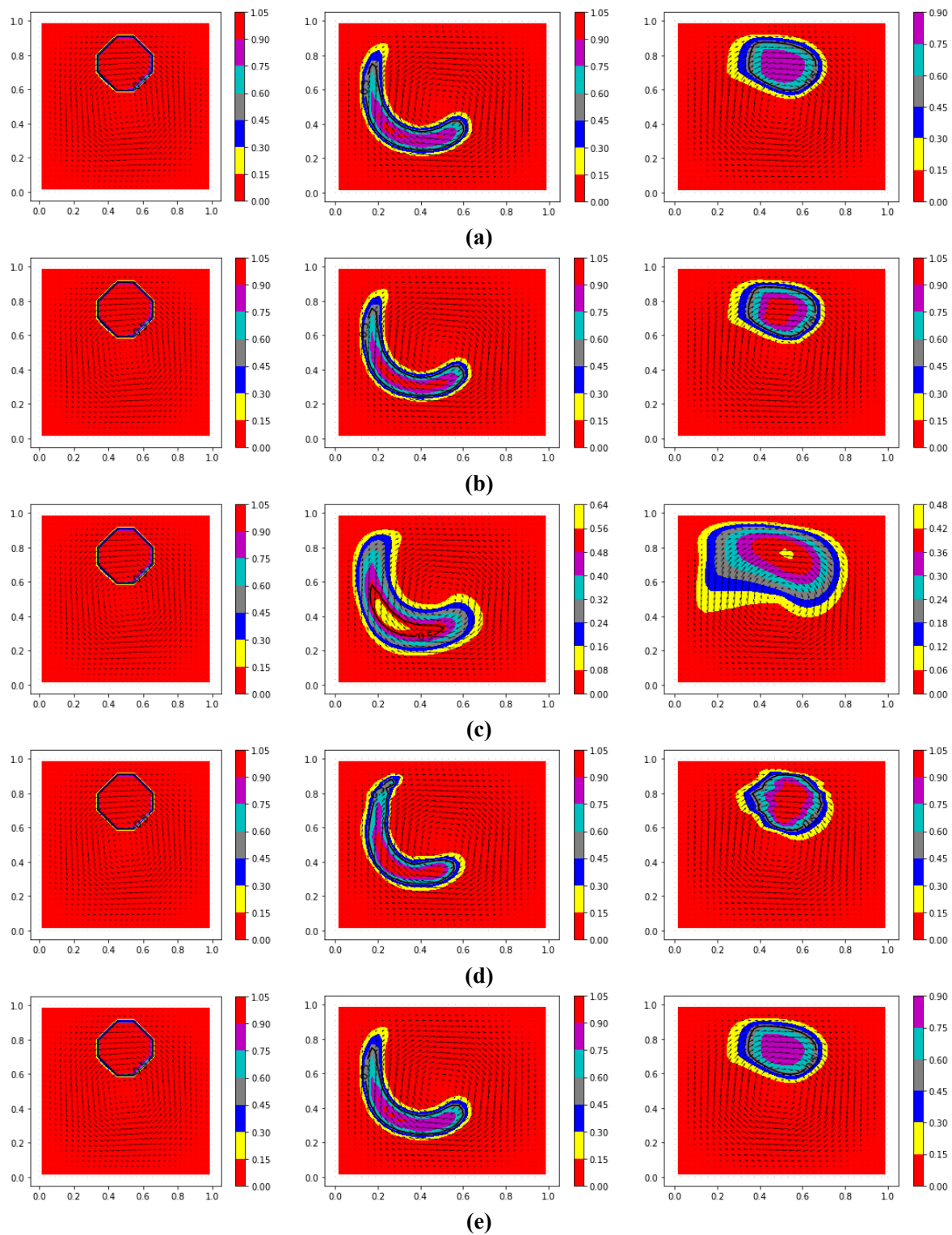

**Figure 8:** The volume fraction contours at t=0 (left), t=T/2 (center), and t=T (right) for a) MULES with IC=0.1 (Multimedia available online)
b) MULES with IC=0.5 (Multimedia available online)
c) first-order upwind
d) second-order central
e) van Leer flux limiter methods. The interface is depicted with a black iso-line of α=0.5.

## 3. Results and Discussion
### 3.1 Effect of Interface Compression (IC)
As mentioned in section 2.4, the IC value is typically fixed between 0 and 2. However, Berberovic et al. and Deshpande et al. expanded the range and examined this constant value up to 4 [15,48]. If the IC value is zero, no compression will occur at the interface. A value between 0 and 1 indicates moderate compression, while values greater than 1 indicate enhanced compression. In this section, IC values between 0 and 2 are considered to evaluate their impact on interface dynamics and mass conservation within the droplet (secondary phase). This investigation focuses on a spherical droplet within a 2D vortex problem, using a grid resolution of $N_x = N_y = 32$ and a time step of $\Delta t=0.005s$, as detailed in section 2.1. Figure 9a-i illustrates the volume fraction contours derived from the MULES method with varying IC values at the final time, $T$.

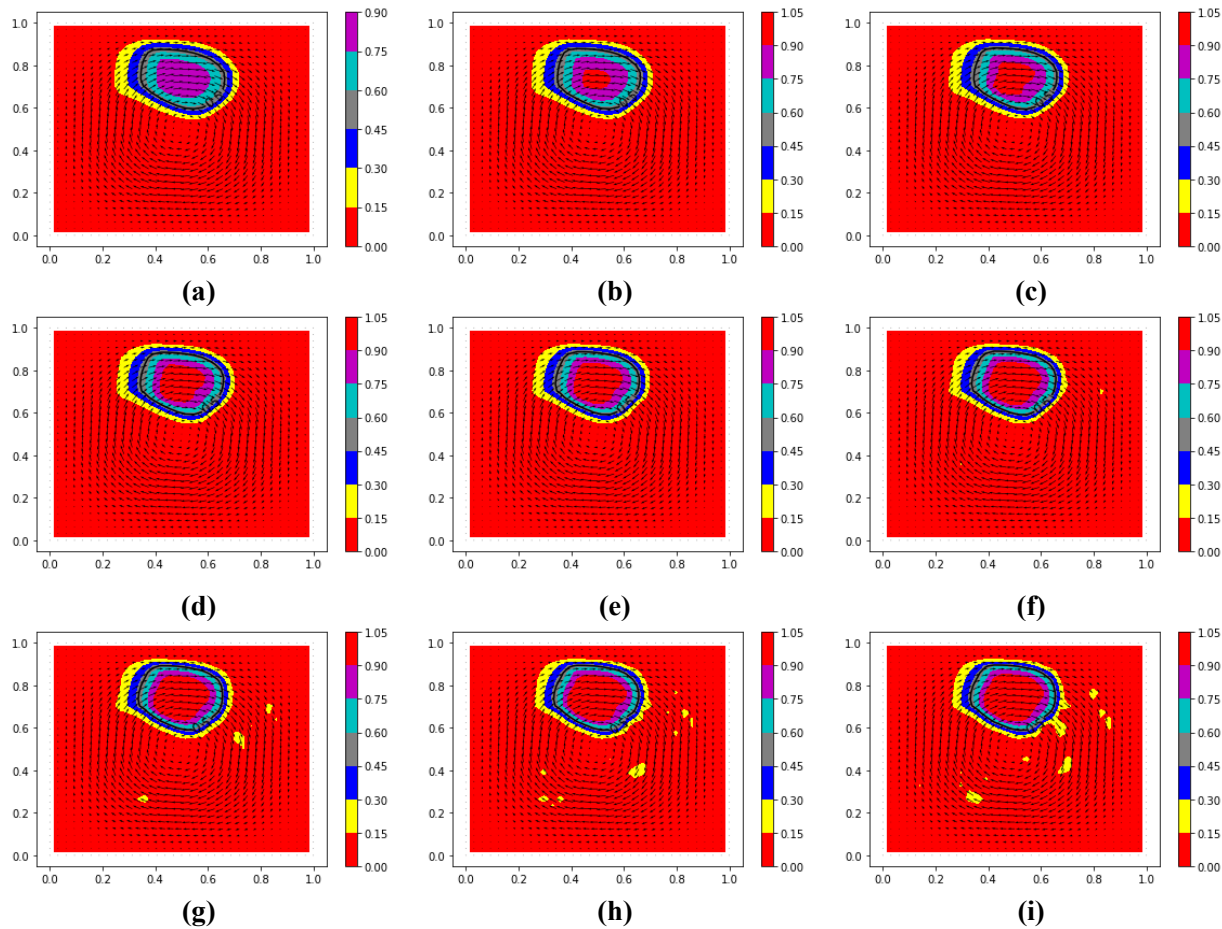

**Figure 9:** The volume fraction contours at t=T for MULES method with a) IC=0, b) IC=0.25, c) IC=0.5, d) IC=0.75, e) IC=1, f) IC=1.25, g) IC=1.5, h) IC=1.75, and i) IC=2. The interface is depicted with a black iso-line of α=0.5.

Figure 9 demonstrates that as the IC value increases, the interface thickness becomes more significant, leading to more accurate interface tracking. For instance, in Figure 9a, when the IC value is zero, there is no phase within the droplet, and the maximum volume fraction inside the droplet is 0.9. However, as the IC value increases, the maximum volume fraction inside the droplet reaches 1, and the phase inside the droplet is considered in the simulation. As the IC value increases, the compressibility of the interface also increases, leading to the volume fraction of 1 inside the droplet advancing towards the interface. However, if the IC value exceeds 1, non-physical instabilities may form during the simulation, disturbing the mass conservation of the droplet. Therefore, it is recommended to keep the IC value between 0 and 1. To substantiate this claim, the results of IAE and MCE are presented in Table 4 and Figure 10. According to the data in Table 4 and Figure 10, the IAE decreases from 4.8% to 3.95% as the IC value increases from 0 to 2. This decline continues steeply until the IC reaches 1.4, where the IAE reaches 3.8%, after which this error fluctuates around 4%. Therefore, it can be concluded that increasing the IC value initially has a favorable effect on interface tracking, but this effect diminishes once the IC surpasses the critical value of 1.4. Conversely, the MCE rises from 0% to 23.19% as the IC increases from 0 to 2. This steep increase in MCE is due to parasitic currents at the interface and instability in the numerical solution. By comparing both errors in Figure 10, it is evident that IAE is almost independent of IC changes, while IC significantly affects the droplet mass conservation. Additionally, the time results reported in Table 4 indicate that changes in IC have no effect on the elapsed time for each simulation. The ET for all simulations remains consistently around 150 seconds.

| Parameter | Interface capturing method: MULES | | | | |
|---|---|---|---|---|---|
| | IC = 0.01 | IC = 0.05 | IC = 0.1 | IC = 0.2 | IC = 0.4 |
| IAE (%) | 4.8 | 4.72 | 4.63 | 4.48 | 4.24 |
| MCE (%) | 0.07 | 0.36 | 0.82 | 1.79 | 3.71 |
| ET (s) | 153.98 | 147.96 | 146.51 | 147.78 | 148.77 |
| | IC = 0.5 | IC = 0.7 | IC = 0.9 | IC = 1 | IC = 1.2 |
| IAE (%) | 4.22 | 4.08 | 3.97 | 3.92 | 3.85 |
| MCE (%) | 5.64 | 8.01 | 10.21 | 11.11 | 13.24 |
| ET (s) | 149.82 | 154.78 | 153.26 | 152.48 | 153.75 |
| | IC = 1.4 | IC = 1.5 | IC = 1.6 | IC = 1.8 | IC = 2 |
| IAE (%) | 3.8 | 4 | 4.09 | 4.04 | 3.95 |
| MCE (%) | 15.37 | 18.95 | 21.37 | 22.96 | 23.19 |
| ET (s) | 152.69 | 152.67 | 153.53 | 158.09 | 151.88 |

**Table 4:** The Comparison Between the Different Values of IC in MULES Scheme, Based on Three Criteria: IAE, MCE, and ET.

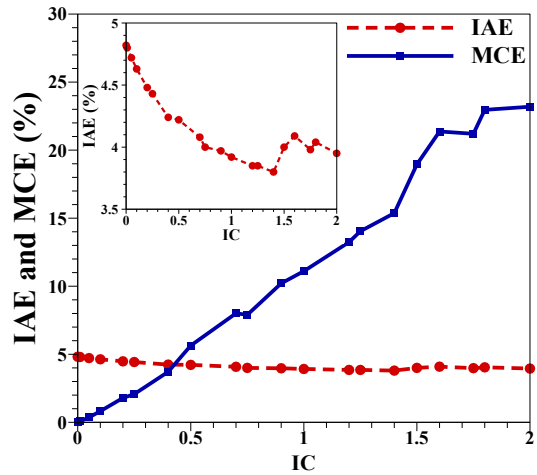

**Figure 10:** The Percentage of Interface Advection Error (IAE) and Mass Conservation Error (MCE) versus the Interface Compression (IC) coefficient. The Inset Provides a Detailed View of the IAE Exclusively.

### 3.2 Effect of Smoothing the Volume Fraction

In this section, surface tension is calculated using the relationships introduced in section 2.4. The impact of volume fraction smoothing and noise reduction on this quantity is examined using the filters outlined in Table 1 (C1 to C4). This investigation focuses on a stationary spherical droplet problem ($x_c = y_c = 0.5$ m and $u = v = 0$ m/s), using a grid resolution of $N_x = N_y = 32$ and a time step of $\Delta t = 0.00125$s. Table 5 and Figure 11a-b present various parameters for both the MULES and van Leer flux limiter methods, including interface advection error, dimensionless pressure difference, capillary number, and elapsed time. Additionally, Figure 12a-e displays the volume fraction contours along with parasitic current for various schemes at final time, $t = 0.5$ s.

| Parameter | Interface capturing method: MULES | | | | |
|---|---|---|---|---|---|
| | without filter | C1 | C2 | C3 | C4 |
| IAE (%) | 0.7 | 0.9 | 0.79 | 0.7 | 0.63 |
| DPD | 1.22957 | 0.99209 | 1.01283 | 1.01614 | 1.01509 |
| Ca | 0.02484 | 0.00663 | 0.00671 | 0.00752 | 0.00868 |
| ET (s) | 136.56 | 143.74 | 145.82 | 147.61 | 150.06 |
| Parameter | Interface capturing method: van Leer flux limiter | | | | |
| | without filter | C1 | C2 | C3 | C4 |
| IAE (%) | 0.69 | 0.9 | 0.79 | 0.7 | 0.63 |
| DPD | 1.22155 | 0.99208 | 1.01281 | 1.01620 | 1.01509 |
| Ca | 0.02472 | 0.00663 | 0.00671 | 0.00753 | 0.00868 |
| ET (s) | 130.01 | 132.21 | 134.90 | 138.46 | 144.18 |

**Table 5:** The Comparison Between the two Interface Capturing Strategies, MULES and Van Leer Flux Limiter, Based on Four Criteria: IAE, DPD, Ca, and ET. (C1 to C4 are introduced in Table 1).

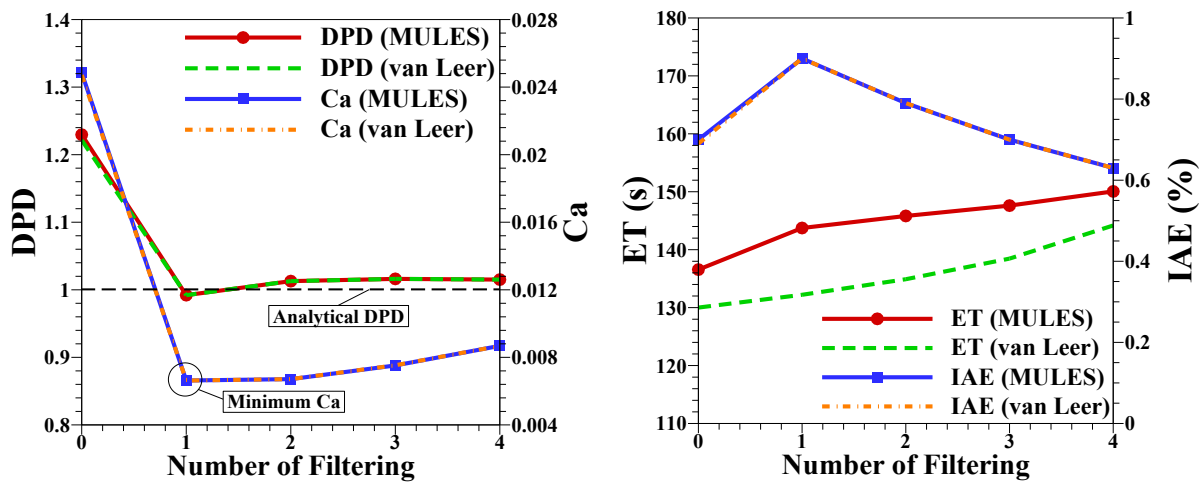

**Figure 11:** The Results of two Interface Capturing Methods, MULES and The Van Leer Flux Limiter, include: a) DPD and Ca, and b) ET and IAE Versus Number of Filtering.

According to the data in Table 5 and Figure 11a-b, both the MULES and van Leer flux limiter methods perform similarly, indicating that volume fraction smoothing has a consistent effect on both schemes of tracking the interface (see Figure 12a-e) (Multimedia available online). A closer examination of Table 5 and Figure 11a reveals that, as the number of filtering steps increases from 0 to 4, DPD and Ca initially decrease. However, after the first filtering, the values of these two quantities increase. With one filtering step, the interface fluctuations are reduced, leading to a decrease in the maximum velocity of the parasitic flow around the droplet by almost 75%, as shown in Figure 12b (Multimedia available online) and Figure 13. Consequently, the capillary number decreases significantly from 0.02484 to its lowest value of 0.00663, representing a 73.31% reduction. A smaller capillary number, closer to zero, indicates higher stability of the solution. Moreover, with one filtering step, DPD reduces by 0.23, approaching the analytical DPD value of 1. However, with further increases in the number of filtering steps, DPD and CA increase, likely due to numerical errors from excessive smoothing of the volume fraction (see Figure 12c-e) (Multimedia available online). Consequently, DPD tends toward an approximate limit value of 1.016, experiencing increases of 0.0162 and 0.01509 compared to the analytical DPD value in the third and fourth filters, respectively. Additionally, according to Figure 13, the maximum parasitic velocity around the droplet increases by 30.92% percent from 0.0663 m/s value to 0.0868 m/s, causing an upward trend in CE from 0.00663 to 0.00868 (also refer to Figure 12c-e) (Multimedia available online). Furthermore, according to Table 5 and Figure 11b, filtering the volume fraction once increases IAE from 0.7 to 0.9. This is likely because minimizing parasitic currents and approaching the real pressure difference inside and outside the droplet makes tracking the interface more challenging, thus increasing the numerical error for both methods. However, with an increased number of filtering steps, IAE decreases from 0.9 to 0.63. As more filtering is performed, fluctuations in the volume fraction are further reduced, resulting in smoother interfaces that the solver can track with less error. It should be noted that filtering imposes additional computational cost, approximately 10 to 15 seconds for the MULES method and 2 to 14 seconds for the van Leer flux limiter method, on the simulation.

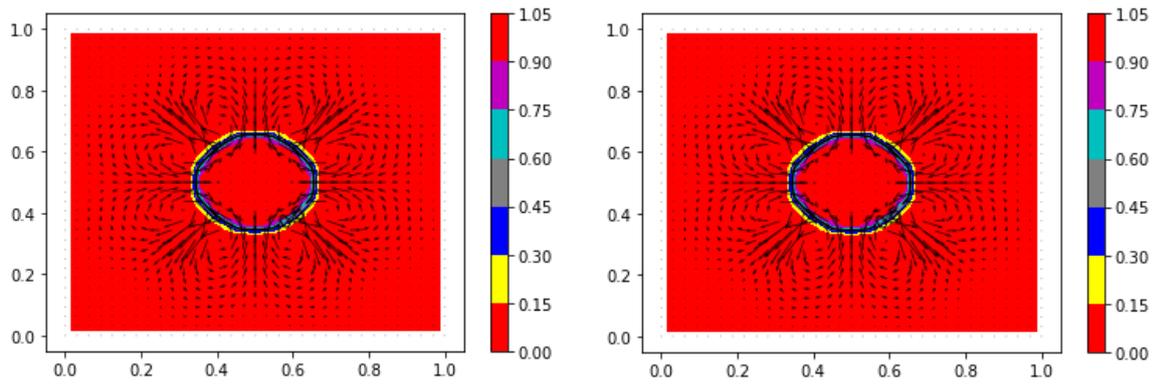

(a)

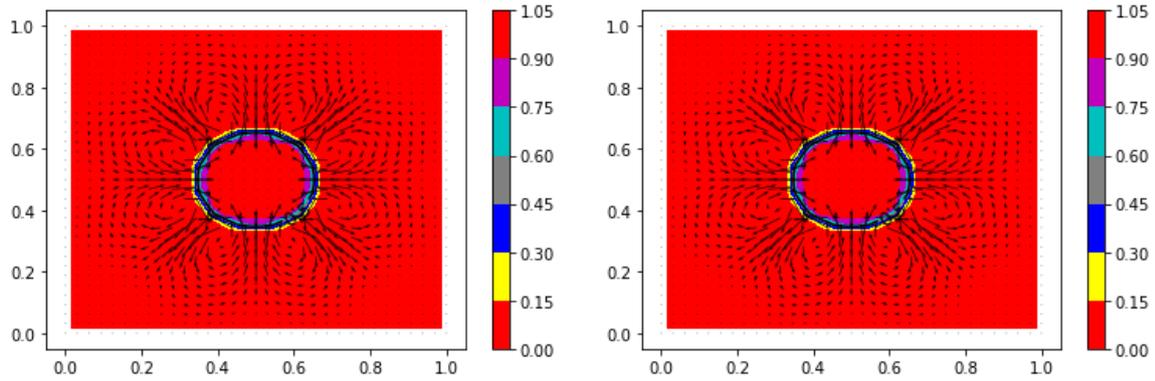

(b)

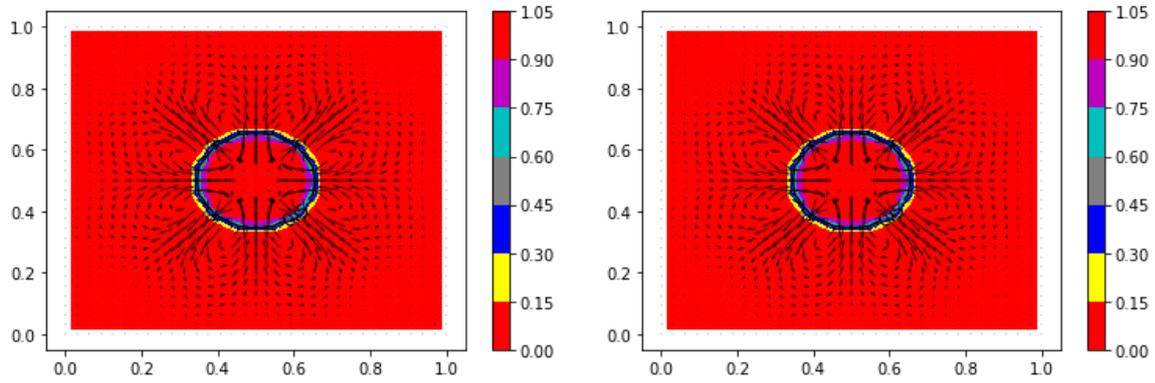

(c)

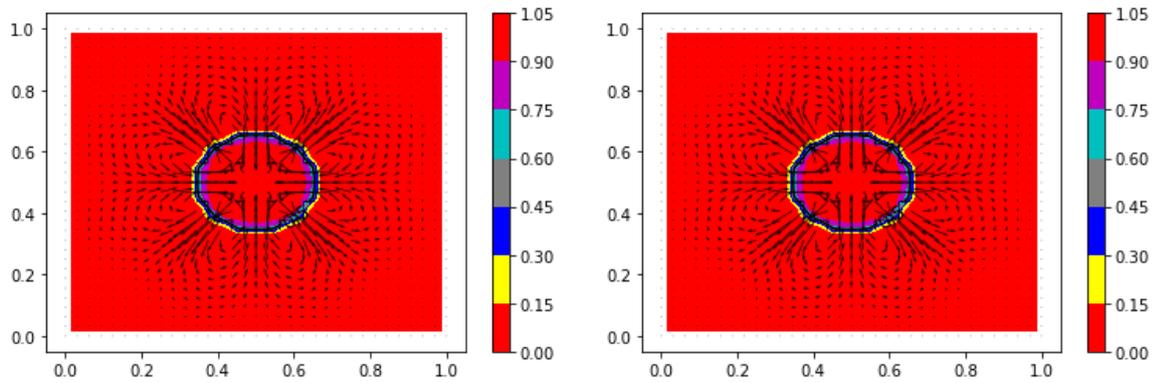

(d)

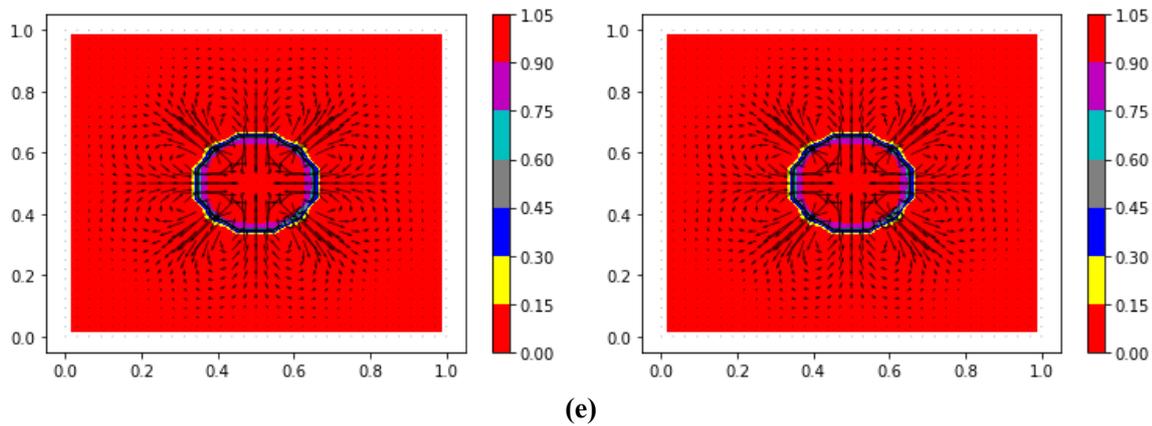

**Figure 12:** The volume fraction contours at final time, t=0.5 s, for MULES (left) and van Leer flux limiter (right) schemes with varying numbers of filtering: a) zero, b) one, c) two, d) three and e) four. The interface is depicted with a black iso-line of α=0.5. (Multimedia available online)

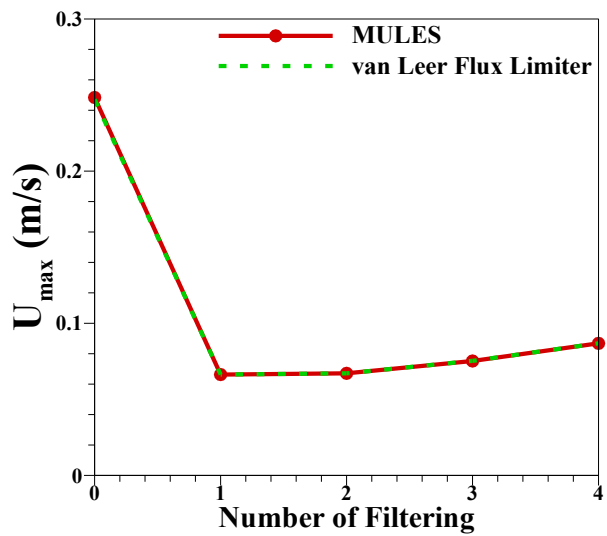

**Figure 13:** The Maximum Velocity of Parasitic Flow Versus Number of Filtering.

## 4. Conclusion

In the present work, an efficient algebraic scheme known as MULES was implemented for sharp interface advection. This method was rigorously verified and compared with several other schemes, including first-order upwind, second-order central, van Leer flux limiter, and Geometric VOF. Two problems involving a spherical droplet in a 2D vortex and a stationary droplet were investigated. The verified model was then employed to examine the effects of the Interface Compression (IC) coefficient, varying from 0 to 2, representing no compression, moderate compression, and enhanced compression. Several key objective parameters, including Interface Advection Error (IAE), Mass Conservation Error (MCE), and Elapsed Time (ET), were analyzed. As the IC value increases, the thickness of the interface becomes more distinct, enhancing the accuracy of the interface tracking within the simulation. At very low IC values, the droplet lacks a discernible phase, and the volume fraction remains limited. However, as the IC value rises, the simulation effectively captures the internal phase of the droplet, leading to a more realistic representation. Interestingly, while the accuracy of interface tracking improves with increasing IC values, this enhancement comes with a caveat. When the IC value surpasses a certain threshold, the simulations begin to exhibit non-physical instabilities, which can disrupt the conservation of mass within the droplet. These instabilities manifest as erratic behaviors in the simulation, particularly in the form of parasitic currents that arise at the interface, compromising the overall integrity of the results. While IAE showed an almost steady improvement as the IC increased, MCE displayed a heightened sensitivity to changes in IC. This disparity highlights the complex relationship between the IC value and mass conservation, underscoring the potential pitfalls of selecting an excessively high IC value. Additionally, MULES and van Leer flux limiter schemes were employed to examine the effects of volume fraction smoothing. Initially, filtering reduces the Dimensionless Pressure Difference (DPD), but beyond the first step, DPD begins to rise again, eventually approaching a value slightly above the analytical prediction. This pattern suggests that while some filtering can stabilize the system, excessive filtering introduces numerical errors that offset the initial benefits. The Capillary Number (Ca) experiences a significant drop after the first filtering step, indicating improved solution stability. However, additional filtering causes the Ca to increase again, reflecting the same trend observed with DPD. The velocity of parasitic flow around the droplet initially decreases markedly with the first filtering step, leading to reduced interface fluctuations and a more stable system. However, with further filtering, the parasitic flow velocity starts to increase again, suggesting that excessive filtering can destabilize the system. The IAE behaves differently compared to other parameters. Initially, IAE increases with one filtering step, likely due to the challenge of accurately tracking a more stable interface. However, as more filtering steps are applied, IAE decreases, indicating that the solver can better track the smoother interfaces produced by extensive filtering.